\documentclass[aps,preprint,showpacs,superscriptaddress]{revtex4}
\date{\today}

\usepackage{graphicx}

\begin{document}

\title{Various Cluster Radioactivities above Magic Nuclei}

\author{F.R. Xu}
\affiliation{School of Physics and MOE Laboratory of Heavy Ion Physics,
Peking University, Beijing 100871, China}
\affiliation{Institute of Theoretical Physics, Chinese Academy of Sciences,
Beijing 100080, China}
\affiliation{Center for Theoretical Nuclear Physics, National Laboratory
for Heavy Ion Physics, Lanzhou 730000, China}
\author{J.C. Pei}
\affiliation{School of Physics and MOE Laboratory of Heavy Ion Physics,
Peking University, Beijing 100871, China}

\begin{abstract}
We present parameter-free tunneling calculations for various cluster 
radioactivities including the diproton decays of atomic nuclei. 
An uniform folded cluster potential has been suggested that is based on 
a self-consistent mean-field model, with the folding factor determined 
using the quantization conditions of the quasibound cluster state. 
We have investigated the $\alpha$-particle and heavier-cluster decays 
of trans-$^{100}$Sn and trans-$^{208}$Pb nuclei, and the observed diproton 
emission from the proton drip-line nucleus $^{16}$Ne, showing the overall 
reasonable descriptions of cluster radioactivities 
with calculated half-lives agreeing well with experimental data. 
We have also predicted the properties of yet unobserved cluster decays 
of the exotic nuclei $^{112,114}$Ba, $^{104}$Te and $^{38}$Ti.
\end{abstract}

\pacs{23.60.+e, 23.70.+j, 21.10.Tg, 27.60.+j, 27.90.+b}

\maketitle 

The charged-particle emission is one of the most important decay modes of 
atomic nuclei. Glancing at the chart of nuclides, one can find that almost all
observed proton-rich exotic nuclei starting from $A\sim 150$ have
$\alpha$ radioactivities. $\alpha$ decays occur more widely in nuclei
heavier than lead isotopes. By a simple picture, the $\alpha$ decay can be
understood as a process of the $\alpha$ particle being formed in the
mother nucleus and emitted tunneling through a Coulomb barrier which is 
created due to the Coulomb interaction between the cluster and 
the remaining nucleons (i.e. the daughter system).
The study of $\alpha$ decays has provided rich information about the
structures of nuclei.

The emissions of heavier clusters, such as $^{14}$C, $^{20}$O, $^{24}$Ne, 
$^{28}$Mg and $^{32}$Si, have been well established experimentally 
in trans-lead nuclei decaying into daughters around the doubly magic 
nucleus of $^{208}$Pb~\cite{Rose84,Pr89,Ho89}.
A second island of heavy-cluster radioactivities 
was predicted~\cite{Gr89} in trans-tin nuclei decaying into daughters 
close to the doubly magic nucleus of $^{100}$Sn. Intense studies, 
particularly on the most promising case of the $^{12}$C emission 
from the proton drip-line nucleus $^{114}$Ba decaying into $^{102}$Sn, 
have been made both experimentally~\cite{Og94,Gu95,Gu97,Ma02} and 
theoretically~\cite{Po93,Ku94,Bu94,Fl95}. 
Experiments have not pinned down the observation of the 
$^{12}$C decay of $^{114}$Ba. Theoretically, the predictions of the partial 
half-life can be different by several orders of 
magnitude~\cite{Po93,Ku94,Bu94,Fl95}. However, the recent 
experiment~\cite{Ma02} has derived the $Q$ value for the possible 
$^{12}$C decay of $^{114}$Ba, which is very important for the theoretical 
prediction because the calculated half-life depends dramatically 
on the $Q$ value. At the proton drip line, another exciting phenomenon is
the diproton radioactivity of nuclei. Modern facilities have been opening 
new perspectives for the study of exotic decays.

The decay process of a charged cluster can be treated with the 
Wentzel-Kramers-Brillouin (WKB) approach. In the present work, 
we have investigated cluster decays from ground states to ground states 
in even-even nuclei. In such decay, the cluster does not carry any 
angular momentum, i.e. $L=0$. The decay width can be written as
(see, e.g.,~\cite{Bu93})
\begin{equation}
\Gamma=PF\frac{\hbar^2}{4\mu}\exp\Big[-2\int^{r_2}_{r_1} dr k(r)\Big],
\label{Gamma}
\end{equation}
where $P$ is the preformation probability of the cluster being formed 
in the mother, and $\mu$ is the reduced mass of the
cluster-daughter system. The normalization factor $F$ is given by
\begin{equation}
F\int^{r_1}_{0}dr\frac{1}{k(r)}\cos^2\Big[\int^r_0 dr^\prime k(r^\prime)
-\frac{\pi}{4}\Big]=1,
\end{equation}
and the wave number $k(r)$ is defined by
\begin{equation}
k(r)=\sqrt{\frac{2\mu}{\hbar^2}|Q-V(r)|},
\end{equation}
where $Q$ is the decay energy and $V(r)$ is the cluster
potential in which the cluster moves. $r_1$ and $r_2$ are the classical inner 
and outer turning points, respectively, obtained by $V(r)=Q$. 
Then the half-life of the cluster decay is obtained 
by $T_{1/2}=\hbar\ln 2/\Gamma$.

In the above method, the determination of the cluster potential $V(r)$ is a key
step for the decay calculation. Phenomenologic potentials by
Buck {\it et al.}~\cite{Bu93} and potentials based on effective
nucleon-nucleon interactions \cite{Oh95,Mo00,Re05} have been successfully
applied to the WKB calculations of the charged-particle decays of nuclei.
The cluster potential can be written as 
\begin{equation}
V(r)=V_{\rm N}(r) + V_{\rm C}(r),
\label{potential}
\end{equation}
where $V_{\rm N}(r)$ and $V_{\rm C}(r)$ are the nuclear potential and
the Coulomb potential, respectively, between the cluster and the daughter 
system. In the present work, the nuclear potential is constructed by
nucleon potentials with multiplying a folding factor $\lambda$ as follows
\begin{equation}
V_{\rm N}(r)=\lambda (N_{\rm c}v_{\rm n}(r) + Z_{\rm c}v_{\rm p}(r)),
\end{equation} 
where $N_{\rm c}$ and $Z_{\rm c}$ are the neutron and proton numbers of 
the cluster, respectively; $v_{\rm n}(r)$ and $v_{\rm p}(r)$ are 
the neutron and proton potentials (excluding the Coulomb potential) 
respectively, obtained from a self-consistent mean-field model. Folding
processes have been employed in other forms of $\alpha$-nucleus potentials 
based on effective nucleon-nucleon interactions~\cite{Oh95,Mo00,Re05}. 
The Coulomb potential $V_{\rm C}(r)$ is well defined physically and 
should not be folded. We have approximated the Coulomb potential by
$V_{\rm C}(r)=Z_{\rm c}v_{\rm c}(r)$, where $v_{\rm c}(r)$ is the 
proton Coulomb potential obtained from the mean-field calculation. 

In the tunneling model, the Bohr-Sommerfeld quantization condition 
defines a relation between the potential $V(r)$ and decay energy $Q$ for 
a quasibound cluster state. We have used the condition to determine the
folding factor $\lambda$ at a given $Q$ value. It will be seen that thus
determined folded $\alpha$-nucleus potential leads to a realistic
volume integral (per nucleon pair) which is obtained from $\alpha$-particle 
scatterings experimentally~\cite{Oh95,Mo00}. The Bohr-Sommerfeld condition 
is given by~\cite{Bu93}
\begin{equation}
\int^{r_1}_0 dr \sqrt{\frac{2\mu}{\hbar^2}|Q-V(r)|}=(2n+1)\frac{\pi}{2}
=(G+1)\frac{\pi}{2},
\end{equation}
where $n$ is the node number in the $L=0$ radial wave function of the 
relative motion of the cluster in the potential, and $G$ is the oscillator
quantum number with $G=2n$~\cite{Bu93}. The quantum number $G$ can be 
determined by the Wildermuth condition, 
$G=\sum^{\rm A_c}_{i=1} g_i$~\cite{Mo00}, where ${\rm A_c}$ is the mass
number of the cluster and $g_i$ is the oscillator 
quantum number of a cluster nucleon orbiting in the mean field.
Actually, $g_i$ is just the main quantum number in the Nilsson 
labelling. For the lowest-energy cluster decay, cluster nucleons should occupy 
orbits immediately above the Fermi levels of the daughter nucleus. 
For trans-$^{100}$Sn nuclei, for example, cluster nucleons occupy 
the $g=4$ orbits, making $G=4{\rm A_c}$. In trans-$^{208}$Pb nuclei, 
the orbits occupied by cluster nucleons have $g=5$ for protons and 
$g=6$ for neutrons, leading to $G=5Z_{\rm c}+6N_{\rm c}$. 
Hence, the combined use of the Bohr-Sommerfeld and Wildermuth quantization
conditions can determine the folding factor $\lambda$. 

Now it has been seen that the present method does not bring any
extra adjustable parameter. The parameter-free calculation is 
particularly useful for exotic nuclei where experimental data are lacking. 
Calculations based on 
self-consistent mean-field models should in principle give more consistent
descriptions on both the structures and decays of nuclei. 
It has been known that mean-field models can in general produce the
structure properties of exotic nuclei~\cite{Be03}. Mean-field potentials,
due to their self-consistent interdependence of proton and neutron
densities that are fed back into potentials, automatically contain 
a dependence on nucleon numbers, i.e. an isospin dependence, in
a self-consistent manner~\cite{Du06}. 

In a very recent work~\cite{Du06}, a form of the
$\alpha$-nucleus potential by $V(r)=2v_{\rm p}(r) + 2v_{\rm n}(r)$
has been used with $v_{\rm p}(r)$ and $v_{\rm n}(r)$ calculated by mean-field
models. However, an extra parameter determined by fitting decay 
data is needed for $\alpha$-decay calculations~\cite{Du06}. Also,
it will be seen that the $\alpha$-nucleus potential without folding 
gives a too deep well, leading to a too large volume integral. 
Hence, the folding process is also to reduce 
the depth of the cluster potential.

In our calculations, a preformation factor of $P=1$ has been assumed for
the various cluster decays of even-even nuclei, as suggested in
Refs.~\cite{Bu93,Bu94}. The potentials $v_{\rm n}(r)$, $v_{\rm p}(r)$ and
$v_{\rm c}(r)$ have been calculated in the spherical daughter system using
the Skyrme-Hartree-Fock (SHF) approach with the SkI4 force
that has been developed with good isospin properties~\cite{Re95}.
Actually, we found that different Skyrme forces 
in general lead to similar results for cluster decay calculations. 
Spherical shapes allow us to perform simple one-dimension tunneling 
calculations, as in Refs.~\cite{Bu93,Oh95,Mo00,Re05,Du06,Na96}.
Compared with the preformation factor, the $Q$ value is a much more crucial
quantity that affects the calculated decay half-life, since the half-life is
exponentially dependent on the $Q$ value. Therefore, we have adopted 
experimental $Q$ values that can be obtained from the measured masses
(binding energies) of mothers, daughters and clusters. 

As a test, we have calculated the $\alpha$-decay property of the spherical 
nucleus $^{212}$Po that has been shown to have a significant $\alpha$-cluster 
structure outside the core of the doubly magic nucleus $^{208}$Pb~\cite{Va92}. 
Our WKB calculation leads to $T_{1/2}^\alpha=89$ ns against the experimental 
half-life of $299\pm 2$ ns~\cite{Au03}. The ratio between the calculated 
and measured half-lives gives a preformation factor of 0.3 agreeing with 
the value of 0.3 calculated by the shell-cluster model~\cite{Va92}. 
The folding factor $\lambda$ is determined to be 0.595. 
Such a folded $\alpha$-nucleus potential gives a rather reasonable volume 
integral of $J_{\rm v}=325$ MeV fm$^3$ which is in a good agreement with the
experimental $J_{\rm v}\approx 300-350$ MeV fm$^3$ for a wide range of
nuclei~\cite{Mo00}, obtained from $\alpha$-particle scattering measurements. 
The $\alpha$-nucleus potential without folding~\cite{Du06} leads to a too
large volume integral of $\sim 550$ MeV fm$^3$. 

With the development of experimental techniques, more and more
exotic nuclei far from the stability are being produced. 
Their decay properties are attracting great interest experimentally and
theoretically. In the present work, we have investigated the cluster decays of 
trans-$^{100}$Sn and trans-$^{208}$Pb nuclei in which $\alpha$-particle and 
heavier-cluster radioactivities have been observed or expected.
Table~\ref{alpha} shows calculated $\alpha$-decay properties.
The calculated half-lives agree with experimental data within a factor of 
$\approx 2$, except $^{114}$Ba for which the experiment of Ref.~\cite{Gu95} 
gave a partial half-life of $T^\alpha_{1/2}\geq 1.2\times 10^2$ s.
As predictions, we have also calculated the $\alpha$-decay half-lives 
of the unknown proton drip-line nuclei $^{104}$Te and $^{112}$Ba, shown in
Table~\ref{alpha}. In the calculations, theoretical $Q$ values given by
M{\" o}ller {\it et al.}~\cite{Moller97} have been adopted. 
However, a change of 1 MeV in the $Q$ value can lead to a variation of 
$\approx 3$ orders of magnitude in the half-life calculation. 
From Table~\ref{alpha}, it can be seen that the folding factors of
the $\alpha$-nucleus potentials are quite stable at $\lambda\sim 0.5$ for 
trans-$^{100}$Sn nuclei and $\lambda\sim 0.6$ for trans-$^{208}$Pb nuclei. 
Resulting volume integrals are $J_{\rm v}\sim 290$ and $\sim 330$ MeV fm$^3$
for the trans-$^{100}$Sn and trans-$^{208}$Pb nuclei, respectively.

\begin{table}
\caption{\label{alpha}Calculated half-lives of the $\alpha$ decays 
in trans-$^{100}$Sn and trans-$^{208}$Pb nuclei, compared with 
experimental half-lives obtained from evaluated data in Ref.~\cite{Au03}. 
For $^{114}$Ba, another experimental half-life \cite{Gu95} 
is given for comparison.}
\begin{ruledtabular}
\begin{tabular}{cccccc}
Emitter & Q \cite{Au03} & G & $\lambda$ & T$_{1/2}^{\rm cal}$ & 
T$_{1/2}^{\rm expt}$ \\
       &  (MeV)        &   &           &   (s)         &    (s)         \\
\hline
$^{104}$Te & 6.12\footnotemark[1] & 16 & 0.529 & 7.3E$-$11 & \\
$^{106}$Te & 4.290 & 16 & 0.521 & 1.4E$-$4 & (7.0$\pm$2.0)E$-$5 \\
$^{108}$Te & 3.445 & 16 & 0.484 & 4.9E+0 & (4.3$\pm$0.2)E+0 \\
$^{110}$Te & 2.723 & 16 & 0.472 & 1.6E+6 & (6.2$\pm$0.3)E+5 \\
$^{110}$Xe & 3.885 & 16 & 0.435 & 2.1E$-$1 & (4.8$\pm$3.0)E$-$1 \\
$^{112}$Xe & 3.330 & 16 & 0.435 & 6.1E+2 & (3.0$\pm$0.9)E+2 \\
$^{112}$Ba & 4.26\footnotemark[1] & 16 & 0.539 & 5.4E$-$2 & \\
$^{114}$Ba & 3.530 & 16 & 0.439 & 7.1E+2& (5.9$\pm$2.6)E+1 \\
 & & & & & $\geq$1.2E+2 \cite{Gu95} \\
$^{222}$Ra & 6.679 & 22 & 0.609 & 2.5E+1 & (3.80$\pm$0.05)E+1 \\
$^{224}$Ra & 5.789 & 22 & 0.611 & 2.9E+5 & (3.16$\pm$0.03)E+5 \\
$^{226}$Ra & 4.871 & 22 & 0.612 & 5.8E+10 & (5.05$\pm$0.02)E+10 \\
$^{228}$Th & 5.520 & 22 & 0.611 & 7.8E+7 & (6.03$\pm$0.01)E+7 \\
$^{230}$Th & 4.770 & 22 & 0.611 & 3.6E+12 & (2.38$\pm$0.01)E+12 \\
$^{232}$U &  5.414 & 22 & 0.609 & 3.2E+9 & (2.17$\pm$0.01)E+9 \\
$^{234}$U  & 4.858 & 22 & 0.609 & 1.1E+13 & (7.75$\pm$0.01)E+12 \\
$^{236}$Pu & 5.867 & 22 & 0.605 & 8.4E+7 & (9.02$\pm$0.02)E+7 \\
$^{238}$Pu & 5.593 & 22 & 0.604 & 2.5E+9 & (2.77$\pm$0.00)E+9 \\
\end{tabular}
\end{ruledtabular}
\footnotetext[1]{Theoretical value from~\cite{Moller97}.}
\end{table}

Heavier-cluster radioactivities in trans-$^{208}$Pb have been observed 
well~\cite{Rose84,Pr89,Ho89}. Theoretical investigations have also been
made~\cite{Bu94,Re04}. Our parameter-free calculations 
for the partial half-lives of the heavy-cluster decays agree with 
observations within one order of magnitude, shown in Table~\ref{cluster}. 
The determined folding factors have similar values to that of corresponding 
$\alpha$-nucleus potentials. For the $^{12}$C radioactivity in $^{114}$Ba,
the recent experiment has derived the decay energy
of $Q=19.00\pm 0.04$ MeV~\cite{Ma02}. With the experimental $Q$ value, 
we have calculated a partial half-life of $T_{1/2}=1.5\times 10^{10}$ s and 
a branching ratio of $b=4.7\times 10^{-8}$ 
($b=\Gamma(^{12}{\rm C})/\Gamma(\alpha)$). The experiment by Guglielmetti
{\it et al.}~\cite{Gu97} has estimated a partial half-life
of $T_{1/2}\geq 1.2\times 10^4$ s and a branching ratio
of $b\leq 3.4\times 10^{-5}$ for the $^{12}$C decay of $^{114}$Ba.
The nucleus $^{112}$Ba has not been known experimentally.
We have used the theoretical decay energy of $Q=23.17$ MeV given by 
M{\"o}ller {\it et al.}~\cite{Moller97} to calculate the partial half-life 
of the $^{12}$C decay of $^{112}$Ba, giving $T_{1/2}=67$ s. Kumar {\it et al.} 
gave another calculated value of $Q=21.46$ MeV~\cite{Ku94}, 
leading to a half-life of $9.5\times 10^4$ s by our calculation.   

\begin{table*}
\caption{\label{cluster}Partial half-lives and branching ratios (relative
to the $\alpha$ decay) of the heavy-cluster decays in trans-$^{208}$Pb and
trans-$^{100}$Sn nuclei. For trans-$^{208}$Pb nuclei, measured partial
half-lives have been compiled in Ref.~\cite{Bu94}, and experimental 
branching ratios and $Q$ values have been obtained from Ref.~\cite{Au03}.
For trans-$^{100}$Sn nuclei, numbers in square brackets indicate 
references from which the adopted values come.} 
\begin{ruledtabular}
\begin{tabular}{cccccccc}
Decay & Q (MeV) & G & $\lambda$ & T$_{1/2}^{\rm cal}$ (s) & $b^{\rm cal}$ &
T$_{1/2}^{\rm expt}$ (s) & $b^{\rm expt}$ \\
\hline
$^{222}$Ra($^{14}$C)&33.05&78&0.540&4.3E+10&5.8E$-$10&(1.01$\pm$0.14)E+11 
& (3.0$\pm$1.0)E$-$10 \\
$^{224}$Ra($^{14}$C)&30.54&78&0.540&2.0E+15&1.5E$-$10&(8.25$\pm$2.22)E+15 
& (4.0$\pm$1.2)E$-$11 \\
$^{226}$Ra($^{14}$C)&28.20&78&0.544&2.0E+20&2.9E$-$10&(2.21$\pm$0.96)E+21 
& (2.6$\pm$0.6)E$-$11 \\
$^{228}$Th($^{20}$O)&44.72&112&0.553&7.1E+20&1.1E$-$13&(5.29$\pm$1.01)E+20 
& (1.1$\pm$0.2)E$-$13 \\
$^{230}$Th($^{24}$Ne)&57.76&132&0.563&1.2E+24&3.0E$-$12&(4.10$\pm$0.95)E+24
& (5.6$\pm$1.0)E$-$13 \\
$^{232}$U($^{24}$Ne)&62.31&134&0.556&7.3E+19&4.4E$-$11&(2.50$\pm$0.30)E+20
& (8.9$\pm$0.7)E$-$12 \\
$^{234}$U($^{28}$Mg)&74.11&154&0.567&7.8E+24&1.4E$-$12&(5.50$\pm$1.00)E+25
& (1.4$\pm$0.3)E$-$13 \\
$^{236}$Pu($^{28}$Mg)&79.67&156&0.561&3.6E+20&2.3E$-$13&4.7E+21
& 2E$-$14 \\
$^{238}$Pu($^{32}$Si)&91.19&176&0.571&3.0E+25&8.3E$-$17&(1.89$\pm$0.68)E+25
& 1.4E$-$16 \\
$^{114}$Ba($^{12}$C)&19.00 \cite{Ma02}&48&0.498&1.5E+10&4.7E$-$8&
$\geq$1.2E+4 \cite{Gu97}&$\leq$3.4E$-$5 \cite{Gu97}\\
$^{112}$Ba($^{12}$C)&21.46 \cite{Ku94}&48&0.499&9.5E+4& 5.7E$-$7& & \\
$^{112}$Ba($^{12}$C)&23.17 \cite{Moller97}&48&0.496& 6.7E+1&8.1E$-$4& & \\
\end{tabular}
\end{ruledtabular}
\end{table*}

Around the proton drip line, the diproton radioactivity is another exciting
challenge in both experiment and theory. Experiments have observed a few
examples of diproton emissions from light proton drip-line nuclei~\cite{Au03}. 
Nazarewicz {\it et al.}~\cite{Na96} have investigated diproton radioactivities 
around the doubly magic nucleus $^{48}$Ni in the framework of the WKB method
with the form of diproton potential by $V_{\rm 2p}(r)=2v_{\rm p}(r)$
(They included the Coulomb potential in $v_{\rm p}(r)$). 
The proton potential $v_{\rm p}(r)$ was
calculated using various mean-field models~\cite{Na96}. 
They compared the depths of potentials, resulting in a modification on 
the term of $|Q-V_{\rm 2p}|$ by multiplying an effective mass of 
$m^*/m<1$~\cite{Na96}. Similarly to cluster decays discussed above,
we have approximated the diproton potential by $V_{\rm 2p}(r)=\lambda\times
2v_{\rm p}(r)+2v_{\rm c}(r)$ (Note that our $v_{\rm p}(r)$ excludes
the Coulomb potential). Again, the half-life calculation of the diproton 
decay is dramatically dependent on the decay energy of $Q_{\rm 2p}$. 

The diproton emission from the proton drip-line nucleus $^{16}$Ne decaying
into the proton magic nucleus $^{14}$O has been observed
with an intensity of 100\%~\cite{Au03}. In this nucleus, the two protons
occupy the $1d_{3/2}$ orbits above the $Z=8$ closed shell, leading to $G=4$.
With the experimental decay energy of $Q_{\rm 2p}=1411\pm 20$ keV~\cite{Au03},  
we obtained a folding factor of $\lambda=0.662$ using the Bohr-Sommerfeld
condition. The calculated half-life of the diproton emission is 
$6.5\times 10^{-20}$ s agreeing with the observed half-life of 
$9\times 10^{-21}$ s~\cite{Au03} within one order of magnitude. 
Another promising candidate for the diproton radioactivity is the 
drip-line nucleus $^{38}$Ti decaying into the proton magic nucleus $^{36}$Ca. 
The experiment has estimated a upper limit of 120 ns for the half-life of 
the diproton decay~\cite{Bl96}. With the evaluated decay energy of
$Q_{\rm 2p}=960\pm 260$ keV from Ref.~\cite{Au03}, we have determined a 
folding factor of $\lambda=0.75$ at $G=6$, resulting in the calculated
diproton half-life of $T_{1/2}=120$ ns. 
If the evaluated error of $\pm 260$ keV in the $Q$ value is considered,
our calculations lead to a range of $T_{1/2}=0.26$ ns to 
$1.7\times 10^6$ ns.

In summary, an uniform folded cluster potential has been suggested for 
the WKB calculations of various cluster decays in atomic nuclei. The folded 
potential is constructed by nucleon potentials obtained from the 
self-consistent Skyrme-Hartree-Fock model, with the folding factor determined 
using the Bohr-Sommerfeld quantization condition combined with the Wildermuth 
condition. This leads to the consistent descriptions of the cluster decay 
and shell structure of a nucleus. The folded $\alpha$-nucleus potential 
gives a reasonable volume integral, agreeing well with the experimental
data obtained from $\alpha$-scattering measurements. No adjustable parameter
has been involved in cluster decay calculations, which is particularly useful 
for exotic nuclei since their properties are largely unknown experimentally. 
We have investigated the $\alpha$-particle and heavier-cluster decays 
of trans-$^{100}$Sn and trans-$^{208}$Pb nuclei, and the diproton emission
from the proton drip-line nucleus $^{16}$Ne above the magic $^{14}$O. 
The calculated half-lives agree with experimental data within   
a factor of $\approx 2$ for $\alpha$ decays, and within one order of 
magnitude for heavier-cluster and diproton decays. 
We have also predicted the half-lives of possible $\alpha$ decays 
in the unknown drip-line nuclei $^{104}$Te and $^{112}$Ba, the expected 
$^{12}$C emission from $^{114,112}$Ba, and the diproton decay of $^{38}$Ti. 
In our calculations, deformation effects on cluster radioactivities
have not been taken into account, which will be discussed in our other works.

We thank P.D. Stevenson for letting us use the SHF code, 
and R. Wyss for useful discussions. This work has been supported by the
Natural Science Foundation of China under Grant Nos. 10525520 and
10475002, the Doctoral Foundation of Chinese Ministry of
Education under Grant No. 20030001088. We also thank the PKU Computer Center 
where numerical calculations have been done.


\begin{thebibliography}{99}
\bibitem{Rose84} H.J. Rose and G.A. Jones, Nature (London) {\bf 307}, 245
(1984).
\bibitem{Pr89} P.B. Price, Annu. Rev. Nucl. Part. Sci. {\bf 39}, 19
(1989).
\bibitem{Ho89} E. Hourani, M. Hussonnois, and D.N. Poenaru, Ann. Phys. 
(Paris) {\bf 14}, 311 (1989).
\bibitem{Gr89} W. Greiner, M. Ivascu, D.N. Poenaru, S. Sandulescu,
in {\it Treatise on Heavy Ion Science}, edited by D.A. Bromley (Plenum,
New York, 1989), Vol.8, p.641.
\bibitem{Og94} Yu. Ts. Oganessian {\it et al.}, Z. Phys. A {\bf 349},
341 (1994).
\bibitem{Gu95} A. Guglielmetti {\it et al.}, Phys. Rev. C {\bf 52}, 740 (1995).
\bibitem{Gu97} A. Guglielmetti {\it et al.}, Phys. Rev. C {\bf 56},
R2912 (1997); 

\bibitem{Ma02} C. Mazzocchi {\it et al.}, Phys. Lett. B {\bf 532}, 29
(2002).
\bibitem{Po93} D.N. Poenaru, W. Greiner, R. Gherghescu, Phys. Rev. C 
{\bf 47}, 2030 (1993).
\bibitem{Ku94} S. Kumar, R.K. Gupta, Phys. Rev. C {\bf 49}, 1922 (1994).
\bibitem{Bu94} B. Buck, A.C. Merchant, S.M. Perez, P. Tripe, J. Phys.
G {\bf 20}, 351 (1994); Phys. Rev. Lett. {\bf 76}, 380 (1996).
\bibitem{Fl95} A. Florescu, A. Insolia, Phys. Rev. C {\bf 52}, 
726 (1995).
\bibitem{Bu93} B. Buck, A.C. Merchant, S.M. Perez, At. Data Nucl. Data
Tables 54, 54 (1993).
\bibitem{Oh95} S. Ohkubo, Phys. Rev. Lett. {\bf 74}, 2176 (1995).
\bibitem{Mo00} P. Mohr, Phys. Rev. C {\bf 61}, 045802 (2000);
               Phys. Rev. C{\bf 73}, 031301(R) (2006).
\bibitem{Re05} C. Xu and Z.Z. Ren, Nucl. Phys. {\bf A760}, 303 (2005);
Phys. Rev. C {\bf 69}, 024614 (2004).
\bibitem{Be03} M. Bender, P.H. Heenen, Rev. Mod. Phys. {\bf 75}, 121 (2003).
\bibitem{Du06} Z.A. Dupr{\'e}, T.J. B{\"u}rvenich, Nucl. Phy. {\bf A 767},
81 (2006).
\bibitem{Re95} P.-G. Reinhard, H. Flocard, Nucl. Phys. A {\bf 584}, 467(1995).
\bibitem{Na96} W. Nazarewicz, J. Dobaczewski, T.R. Werner, J.A. Maruhn,
P.-G. Reinhard, K. Rutz, C.R. Chinn, A.S. Umar, M.R. Strayer, Phys. Rev. C
{\bf 53}, 740 (1996). 
\bibitem{Va92} K. Varga, R.G. Lovas, R.J. Liotta, Phys. Rev. Lett. {\bf 69},
37 (1992).
\bibitem{Au03} G. Audi, O. Bersillon, J. Blachol, A.H. Wapstra, Nucl. Phys.
{\bf A 729}, 3 (2003); G. Audi, A.H. Wapstra, C. Thibault, Nucl. Phys.
{\bf A 729}, 337 (2003).
\bibitem{Moller97} P. M{\"o}ller, J.R. Nix, K.-L. Kratz, Atom. Data and Nucl.
Data Tables {\bf 66}, 131 (1997).
\bibitem{Re04} Z.Z. Ren, C. Xu, Z.J. Wang, Phys. Rev. C {\bf 70}, 
034304 (2004).
\bibitem{Bl96} B. Blank {\it et al.}, Phys. Rev. Lett. {\bf 77}, 2893 (1996).

\end{thebibliography}
\end{document}